# Single domain wall effect on parametric processes via Cherenkov-type phase matching

Xuewei Deng, Huaijin Ren, Yuanlin Zheng and Xianfeng Chen*

*Department of Physics, the State Key Laboratory on Fiber Optic Local Area Communication Networks and Advanced Optical Communication Systems, Shanghai Jiao Tong University, Shanghai 200240, China*

*xfchen@sjtu.edu.cn

We report on important influence of single domain wall (DW) of electrically poled ferroelectric crystal on parametric processes via Cherenkov-type phase matching. It shows that the effective nonlinear polarization is confined in DW and its phase velocity can be modulated when incident light is off domain wall's direction. These effects lead to novel Cherenkov second harmonic generation (CSHG) which has no analogue in bulk ferroelectrics. Complex DW-modulated parametric process via Cherenkov-type phase matching, such as Cherenkov sum frequency generation (CSFG), is also demonstrated in DW. We propose that nonlinear coefficient in DW is significantly enhanced.

**OCIS Code:** 190.0190, 190.2620, 240.1485

A charged particle moving faster than light in medium can drive surrounding atoms to emit coherent light called Cherenkov radiation (CR)[1]. CR is always observed in a cone defined by Cherenkov angle $\theta_C$=arc cos(v'/v), where v is the velocity of the moving charged particle and v' is the phase velocity of CR. It's seen that CR occurs only when v>v'. A similar phenomenon called CSHG is observed in nonlinear optics[2-4] when phase of nonlinear polarization propagates faster than phase velocity of free light of the same frequency in medium. While these two processes are similar, CSHG has unique distinguishing features. For example, there is no charged particle but a spatially extended nonlinear polarization driven by the incident light field. Most importantly, CSHG is a nonlinear optical process associated with the susceptibility tensor $\chi^{(2)}$. Typical setup for CSHG is waveguide structure with nonlinear substrate[3]. Recently years, CSHG has been reported in SBN crystal[5-6] and periodically poled ferroelectrics[7-8], which samples have DWs in them. This is a new feature of DW among those newly reported ones, like optical birefringence, strain, local electric fields and electromechanical contrast[9]. Although CSHG in DW is attributed to newly generated nonlinear coefficient $d_{11}$ and $d_{12}$ by strain[7] and CSFG can be used for pulse characterization[10], its detailed process is still not clear so far. What we know from earlier studies[7-8] is only the DW that plays an important role in such processes, but how a single DW works and its influence on the CSHG has not been studied yet.

In this letter, we study parametric processes via Cherenkov-type phase matching in single DW theoretically and experimentally. We find that the nonlinear polarization, which generates the observable CSHG, is confined in DW. By changing the angle of incident light with respect to DW, we can easily change the phase velocity of the nonlinear polarization and consequently change the Cherenkov angle of the CSHG. Different from the reason of newly generated nonlinear coefficient $d_{11}$and $d_{12}$ in previous studies[7,10], by solving Maxwell's equations, we find that the confinement of nonlinear polarization is due to the $10^3$ times significant enhancement of nonlinear coefficient in DW. As another evidence, CSFG modulated by DW is also demonstrated in our experiment, which, to the best of our knowledge, is reported for the first time.

CSHG has been reported in periodically poled ferroelectrics with multiple DWs[7-8]. In order to investigate the influence of DW on the CSHG, we focused our study on CSHG in single DW. Single DW with high quality is fabricated in 1mm Z-cut $LiNbO_3$ sample as shown in Fig. 1. A Y-polarized incident light of 100 fs pulses centered at 800nm with averaged power 200mW is loosely focused (f=15cm) in the DW along z direction with spot size about 100μm. It is seen from the screen as shown in Fig. 1 that a pair of well collimated CSHGs has been generated and they are symmetrical to the DW. In comparison, we cannot observe any CSHG in bulk $LiNbO_3$ with homogenous domain. This experiment demonstrates that CSHGs in periodically poled ferroelectrics indeed originate in their DWs and doesn't show evidence with the periodic $\chi^{(2)}$ structures.

A more important phenomenon in this experiment is the well-collimated CSHGs which are symmetrical to the DW. Let's consider the situations of CSHG in planar waveguide and bulk ferroelectrics. In planar waveguide, the radiation source (second order nonlinear polarization) and incident light (guide modes of the waveguide) have the same phase velocity v. In normal dispersive medium, v is always larger than v' of free second harmonic (SH) beam, and consequently the harmonic radiation emits into the substrate with certain angle.

Since the geometry is planar and the nonlinear polarization is coherent over the full beam width, CSHG emerges as well-collimated beam at Cherenkov angle[3]. Similarly, in bulk ferroelectrics conical CSHG[2] is generated due to the isotropy of the nonlinear polarization forced by incident light. Since the patterns of CSHG are determined by the geometry of the nonlinear polarization, we find that nonlinear polarization in our experiment has planar geometry according to the well-collimated CSHG beams and this planar nonlinear polarization is well confined in the DW because the CSHGs are symmetrical to it. Typical DW width of ferroelectrics is believed to be less than 100nm[11]. It is well explained that the geometry of the CSHG source is planar when compared with the 100μm incident light spot. =

Energy and momentum conservations are both satisfied in this CSHG process in DW. Energy conservation is fulfilled by converting two photons with frequency ω to one photon with frequency 2ω. Momentum conservation is fulfilled by Cherenkov-type phase matching condition, to be discussed in detail below. See in Fig. 2 (a), when incident light is along the direction of DW, it drives second order nonlinear polarization in DW to propagate with it at the same phase velocity $v_{np}$=v. Free SH propagates at a slower phase velocity v' in the same medium. Cherenkov angle is then determined by $\theta_C$=arc cos(v'/v). We change v and v' into momentum space and get the Cherenkov-type phase matching condition as,

$$\cos\theta_C = \frac{2|\vec{k}|}{|\vec{k}'|} \tag{1}$$

where $\vec{k}$ and $\vec{k}'$ are the wave vectors of incident light and SH in the medium respectively. The triangle relationship of these two wave vectors is also shown in Fig. 2 (a). This is typical Cherenkov-type phase matching. However, since the nonlinear polarization is confined in DW, it is easy to find that new interesting CSHG can be generated. As shown in Fig. 2 (b), when we rotate the incident light in X-Z plane so that it has an angle $\gamma$ to Z-axis, the forced nonlinear polarization in DW will propagate with the incident light noncollinearly. At time $t_0$, the equiphase plane of incident light drives nonlinear polarization in DW at point A. With the propagation of this equiphase plane at phase velocity v, the nonlinear polarization will propagate in DW to point B at time t. The phase velocity of the nonlinear polarization has been increased obviously, to be $v_{np}=v/\cos\gamma$. Free SH still propagates at phase velocity v'. So the Cherenkov angle becomes $\theta_C=\arccos(v'\cos\gamma/v)$. We change this relationship to momentum space again and get new phase matching condition for incident light with arbitrary incident angle $\gamma$ with respect to DW in x-z plane,

$$\cos\theta_C = \frac{2|\vec{k}|\cos\gamma}{|\vec{k}'|} \quad . \tag{2}$$

Experimentally, we rotate the sample with single DW in X-Z plane and get the result that the CSHG angle increases with the incident angle, as shown in Fig. 3. The measured external CSHG angle $\theta_c$ changing with different incident angle $\gamma$ is in good agreement with the theoretical calculations by equation (2). By changing incident angle, it is easy to change the phase velocity of the nonlinear polarization from v to infinite. So the CSHG can be modulated continuously. From our experiment, we can conclude that single DW plays a key role in this process.

As discussed earlier, CSHG can always be generated in normal dispersive nonlinear medium. Why it has been enhanced in DW? From Maxwell's equations, we derive the inhomogeneous vector wave equation for CSHG

$$\nabla \times \nabla \times \vec{E} - \frac{n}{c^2}\varepsilon \vec{E} = \omega_h^2 \vec{P}, \qquad (3)$$

where $\omega_h$ is the frequency of the SH, ε is the dielectric constant, $\vec{P}$ is the second order nonlinear polarization. Detailed solving process is discussed in previous study[12] and the solution for CSHG at distance r is

$$\vec{E}(r) = -\frac{n}{r\,4\pi\varepsilon_0 c^2}\vec{P}VX(n,n')D(n,n')e^{-i\frac{\omega_h}{c}n'r} \qquad (4)$$

where n and n' are the refractive indices of incident light and SH respectively, X(n,n') and D(n, n') are functions of n and n' and they equal to 1 in the limit of small radiating source, V is a constant. From equation (4), we find the only reason for the enhancement of CSHG in DW is the enhancement of $\vec{P}$. Since $\vec{P} = d_{eff}\vec{E}\vec{E}$, $d_{eff}$ in DW must have been enhanced. In our previous study[13], the CSHG efficientcy in PPLN is measured to be about 1%, several orders of magnitude greater than that of about $10^{-10}$ in bulk $LiNbO_3$ with homogenous domain[2]. Eliminating the contribution from multiple DWs in PPLN, we estimate the enhancement of $d_{eff}$ in DW is $10^3$. Consequently, enhanced nonlinear coefficient leads to enhanced nonlinear polarization confined in DW, which emits observable CSHG.

By the new effect of DW, complex parametric process via Cherenkov-type phase matching from more incident light beams is available. When two incident light beams, $\omega_1$ and $\omega_2$ are injected at $\gamma_1$ and $\gamma_2$ with respect to DW in X-Z plane, each of them can generate a pair of CSHGs at the angles satisfying equation (2). If they just overlap in DW, a third kind nonlinear polarization $\vec{P}_S = d_{eff}\vec{E}_1\vec{E}_2$ will be generated and it will emit CSFG at a unique angle that is defined by

$$\cos\theta_C = \frac{|\vec{k}_{\omega_1}|\cos\gamma_1 + |\vec{k}_{\omega_2}|\cos\gamma_2}{|\vec{k}_{\omega=\omega_1+\omega_2}|}. \quad (5)$$

This equation is just the definition of CSFG modulated by DW and the momentum conservation is fulfilled by Cherenkov-type phase matching condition. If $\gamma_1$ and $\gamma_2$ equal to 0, equation (5) becomes the definition of typical CSFG[14]. And additionally, if $\omega_1=\omega_2$, it just becomes the definition of CSHG, the same as equation (1).

Experimentally, we split the incident pulse to two identical pulses and tune the delay of one to make them overlap precisely in DW at different incident angles (shown in Fig. 4 (a)). Three pairs of SH beams are observed in our experiment. One side of the SH pattern is shown in Fig. 4. In Fig. 4 (b), the outer and the inner SH spots are DW-modulated CSHGs generated by the incident beams respectively. The middle one is DW-modulated CSFG by both of these two incident lights. If we tune the delay of one pulse and make the two not overlap, the middle one disappears at once (Fig. 4 (c)). No matter either of the two incident pulses is blocked, the corresponding CSHG and the CSFG disappear at once (Fig. 4 (d) and (e)). If we rotate the sample to change the incident angles of the two pulses, the SH spots move continuously and they just overlap like one pair when the incident angles of the two pulses are identical. The existence of DW poses additional confinement on the propagation of effective nonlinear polarization without changing the incident pulses, so it can modulate the nonlinear polarization continuously. It can be expected that DW effect has indeed imposed important influences on conventional nonlinear processes. Making clear understanding of DW effect becomes more and more important.

In conclusion, we demonstrate the important influence of DWs on parametric processes via Cherenkov-type phase matching condition theoretically and experimentally. DW-modulated CSHG and CSFG are observed in our experiment and analyzed. We find the nonlinear polarization forced by incident light is confined in DW and through solving the Maxwell's equations we find the reason is the enhanced nonlinear coefficient in DW. This new characteristic of DW should draw more attentions and more intensive studies are demanded.

This research was supported by the National Natural Science Foundation of China (No. 60508015 and No.10574092), the National Basic Research Program "973" of China (2006CB806000), and the Shanghai Leading Academic Discipline Project (B201).

List of figure captions

Fig. 1 Pattern of CSHG generated in DW. In the left part, the red line represents the incident light and the dashed area is domain inversion fabricated by poling technique. LN means lithiun niobate. The right picture is the DW observed in microscopy.

Fig. 2 Schematic depiction of CSHG processes and the corresponding phase matching condition when (a) the incident light is along the direction of DW; and (b) the incident light has an angle with respect to the DW in x-z plane. EP and NP mean equiphase plane and nonlinear polarization respectively. The red area and the bold large arrow represent the incident beam and its direction.

Fig. 3 External Cherenkov angle changes with the incident angle.

Fig. 4 (a) Schematic of the setup used for the DW-modulated CSFG; (b) shows the observed CSHGs and CSFG when the two incident pulses overlap in DW; (c) is the situation when the two incident pulses do not overlap in DW; (d) and (e) show the results when one of the incident pulses is blocked.

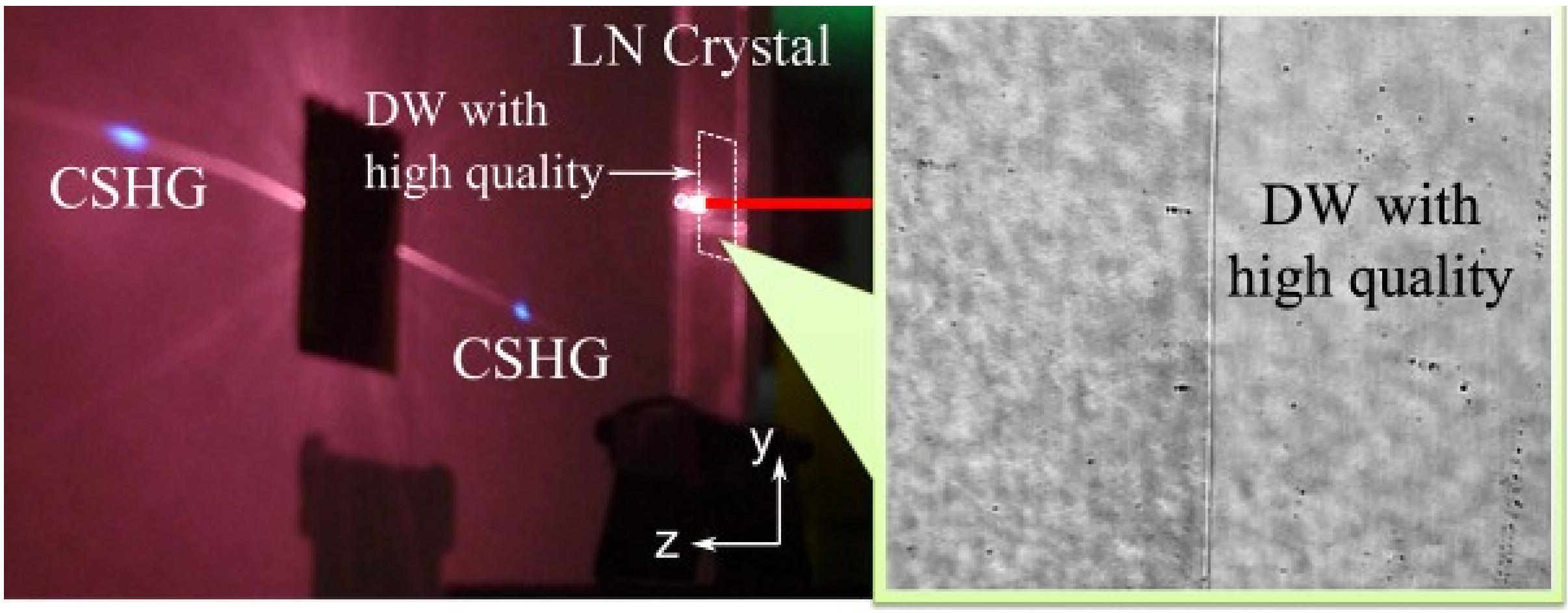

Fig. 1 Pattern of CSHG generated in DW. In the left part, the red line represents the incident light and the dashed area is domain inversion fabricated by poling technique. LN means lithiun niobate. The right picture is the DW observed in microscopy.

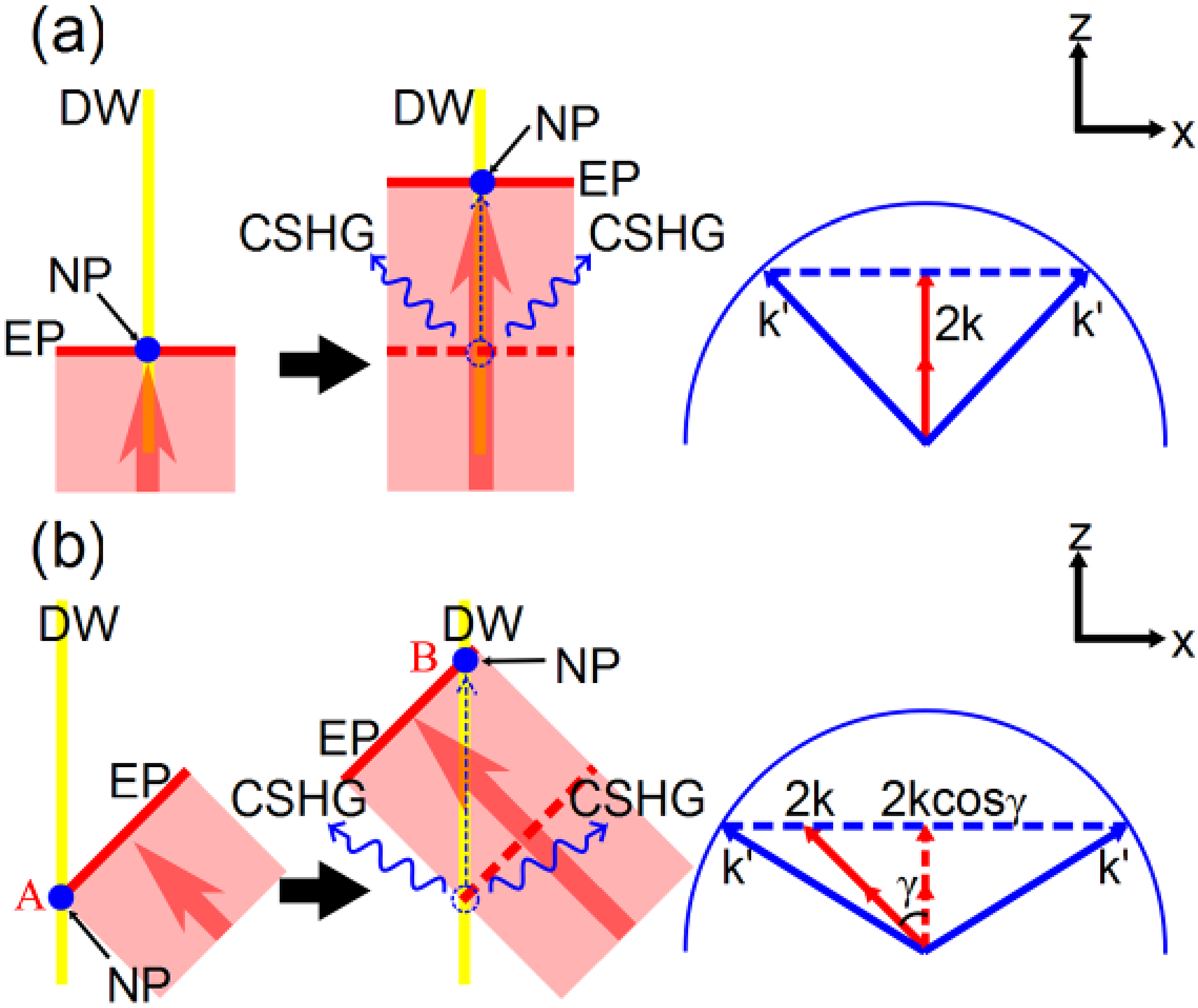

Fig. 2 Schematic depiction of CSHG processes and the corresponding phase matching condition when (a) the incident light is along the direction of DW; and (b) the incident light has an angle with respect to the DW in x-z plane. EP and NP mean equiphase plane and nonlinear polarization respectively. The red area and the bold large arrow represent the incident beam and its direction.

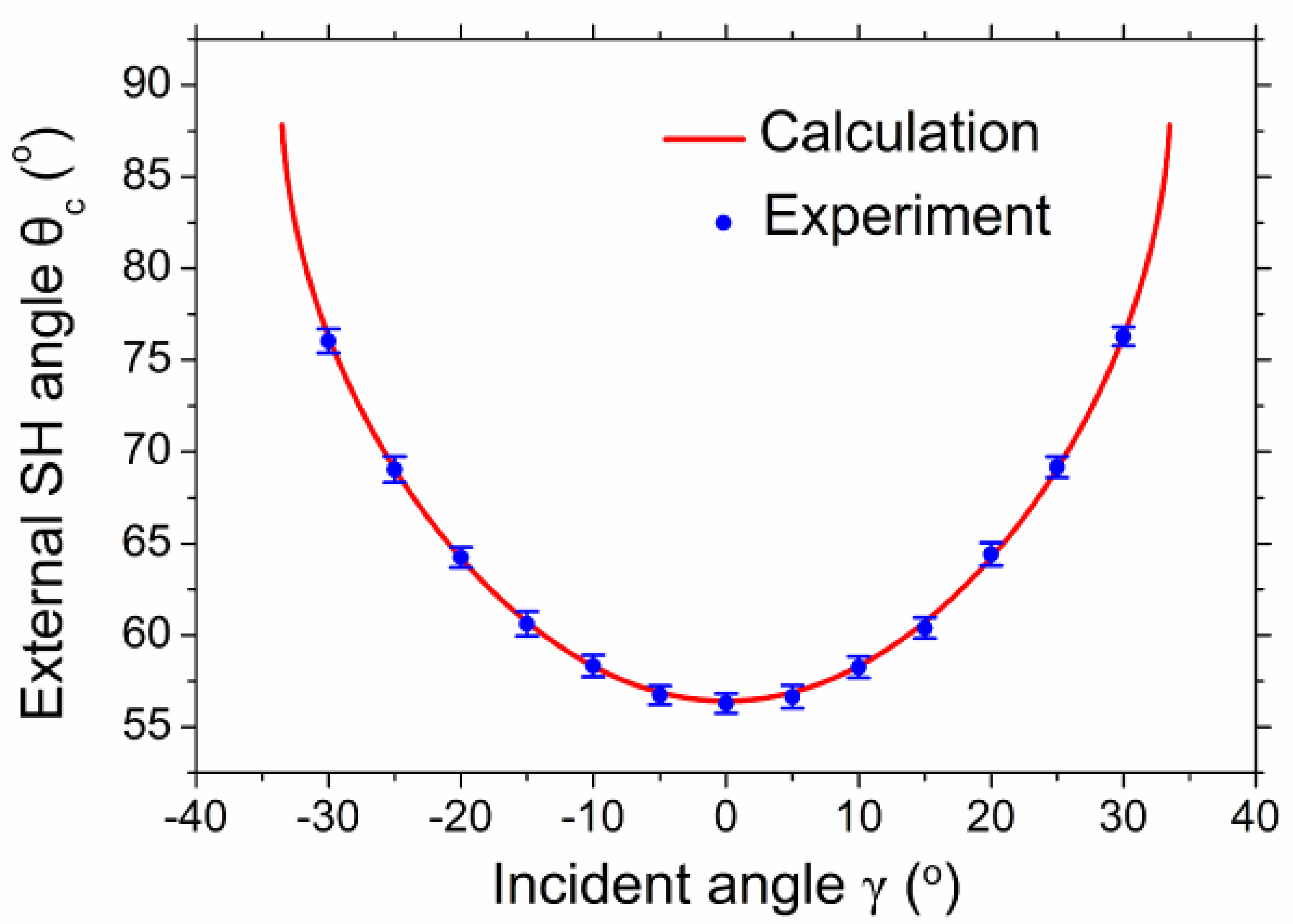

Fig. 3 External Cherenkov angle of CSHG changes with the incident angle.

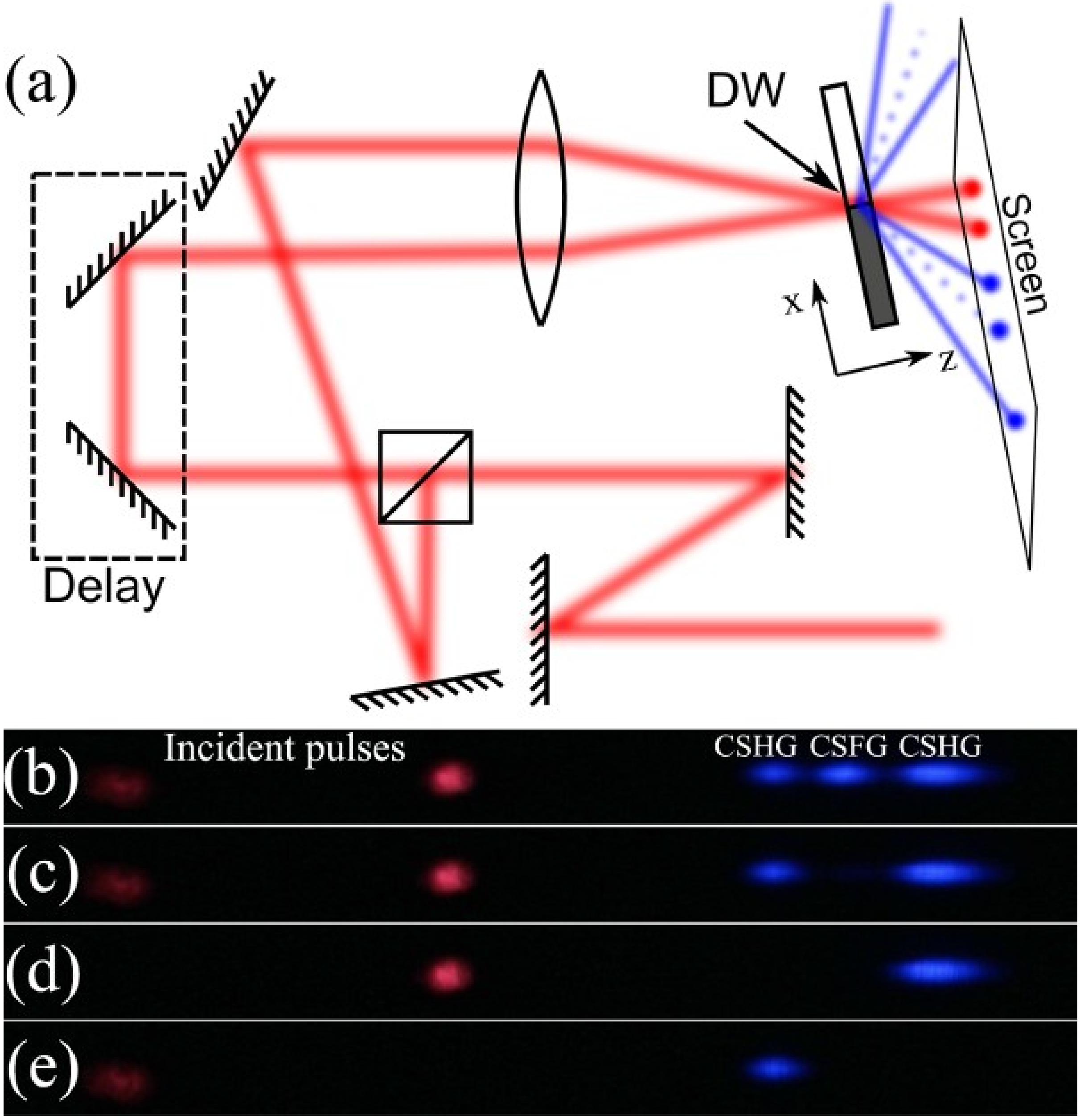

Fig. 4 (a) Schematic of the setup used for the DW-modulated CSFG; (b) shows the observed CSHGs and CSFG when the two incident pulses overlap in DW; (c) is the situation when the two incident pulses do not overlap in DW; (d) and (e) show the results when one of the incident pulses is blocked.